\documentclass{ws-procs975x65}

\def\beq{\begin{equation}}
\def\eeq{\end{equation}}

\begin{document}

\title{Semi-classical and nonlinear energy conditions}
\author{Prado Mart\'{\i}n-Moruno}

\address{Departamento de F\'{\i}sica Te\'orica I, Universidad Complutense de Madrid,\\ E-28040 Madrid,
Spain.
}

\author{Matt Visser} 

\address{School of Mathematics and Statistics,
Victoria University of Wellington,\\
PO Box 600, Wellington 6140, New Zealand.
}

\begin{abstract}
We consider the characteristics of nonlinear energy conditions 
and of quantum extensions of these and the usual energy conditions.
We show that they are satisfied by some quantum vacuum states that violate the usual energy
conditions.
\end{abstract}


\bodymatter

\section{Usual energy conditions}

The energy conditions (ECs) are assumptions made on the matter content to extract generic characteristics of the spacetime geometry from the dynamical equations.
The strong energy condition (SEC) assumes a reasonable behaviour of matter in a given theory of 
gravity, stating that gravity is always attractive in general relativity. 
Thus, it combines the time-like convergence condition with the Einstein
equations leading to:\cite{HyE}
\begin{equation}
  {\rm SEC:}\,\,\, T^{a}{}_b\,V_a\,V^b-\frac{1}{2}T\, V^aV_a\geq0,                                              
\end{equation}
where $V^a$ is any time-like vector.
Nowadays this condition is almost completely abandoned since it is known to be violated 
in cosmological scenarios during the early inflationary phase and the 
current epoch of accelerated expansion of our Universe.

On the other hand, a more fundamental approach consists of assuming properties that any material content should satisfy 
independently of the theory of gravity.
The dominant energy condition (DEC) states that the energy density measured by any observer is non-negative and 
propagates in a causal way. This is
\begin{equation}
 {\rm DEC:}\,\,\, T^a{}_b\,V_a\, V^b\geq0\,\,\&\,\,F^aF_a\leq0\,\,\,{\rm with}\,\,\,  F^a=-T^{a}{}_bV^b,
\end{equation}
for any time-like vector $V^b$.
The weak energy condition (WEC) only states that the energy density measured by any observer has to be non-negative, leading to
\begin{equation}
 {\rm WEC:}\,\,\,T^a{}_b\,V_a\, V^b\geq0.
\end{equation}
The null energy condition (NEC) is a particular limit of both the SEC and the WEC.
For any $k^a$ null-vector, it requires 
\begin{equation}
  {\rm NEC:}\,\,\, T^a{}_b\,k_a\, k^b\geq0.
\end{equation}
These ECs are known to be violated in some semi-classical situations,
as by the renormalized stress-energy tensor of the Casimir vacuum and 
in the Schwarzschild spacetime 
(even unbounded violations for the Boulware and Unruh vacua\cite{B,U}).

\section{New Energy Conditions}

An interesting approach that has been followed in the literature consists in averaging the usual ECs along casual curves curves,
or in constraining the violations of the ECs with some quantum inequalities.\cite{Ford,FordRoman}
Here we consider, however, a local approach.

\subsection{Nonlinear energy conditions}

The flux energy condition (FEC) was first considered in Ref.\citenum{FEC} to obtain some entropy bounds for uncollapsed systems.
The FEC states that the energy density measured by any observer propagates in a causal way, without assuming anything about
the sign of the energy density. That is\cite{FEC,QFEC}
\begin{equation}
{\rm FEC:}\,\,\, F^aF_a\leq0.
\end{equation}
By their very definition, the WEC plus the FEC is equal to the DEC. 
Other nonlinear energy conditions can be formulated\cite{nonlinear} although their physical
interpretation is not so clear.
These are the determinant energy condition (DETEC) and
the trace-of-square energy condition (TOSEC), which state\cite{nonlinear}
\begin{equation}
 {\rm DETEC:}\,\,\, \det \left( T^{ab}\right) \geq 0,\qquad
 {\rm and}\qquad
  {\rm TOSEC:}\,\,\, T^{ab}\, T_{ab} \geq 0.
\end{equation}

\subsection{Semi-classical energy conditions}

One can note that measured violations of the ECs are typically small and associated to semi-classical effects.
Thus, one can formulate some semi-classical ECs based on constraining the violations of the ECs for quantum
states locally.

The quantum FEC (QFEC) states that the energy density measured by any observer either propagates in a causal way, or does not propagate ``too badly'' in an acausal way. Thus, the flux 4-vector can be (somewhat) space-like in semi-classical situations but its norm has to be bounded from above.
Assuming that this bound depends on the characteristics of the system, the QFEC can be formulated as\cite{QFEC}
\begin{equation}
{\rm QFEC:}\,\,\, F^aF_a \leq \zeta \;(\hbar N/L^4)^2\; (U_a\,V^a)^2,
\end{equation}
where $\zeta$ is a parameter of order unity,
$N$ the number of fields, $U^a$ the system 4-velocity, and $L$ a characteristic distance.

Similar quantum generalizations of the other ECs can also be considered.\cite{nonlinear}
The quantum weak energy condition (QWEC) states that the energy density measured by any observer should not be ``excessively negative'', 
this is \cite{nonlinear}
\begin{equation}
{\rm QWEC:}\,\,\,T^{a}{}_b\, V_a V^b \geq -\zeta\hbar N/L^4\; (U_a\,V^a)^2.
\end{equation}
Violations of the QWEC in situations where the QFEC is satisfied would point to the fact that the fulfillment of 
QFEC is not trivially due to the quantum extension.

\subsection{Conditions for different stress-energy tensors}
Without loss of generalitiy, any stress-energy tensor (SET), 4-velocity and associated 4-flux,  can be written in an orthonormal basis as
\beq\label{st}
T^{\widehat{a}\widehat{b}} =
 \left[\begin{array}{c|ccc}\rho&f_1&f_2&f_3\\ \hline f_1&p_1&0&0\\f_2&0&p_2&0\\f_3&0&0&p_3\end{array}\right],
 \qquad V^{\widehat{a}} = \gamma\left(1;\beta_i\right),
 \qquad F^{\widehat{a}} = \gamma\big(\rho-\vec \beta\cdot\vec f ; \;  f_i-p_i\beta_i\big),
\eeq
with  $\sum_i\beta_i^2<1$. The Hawking--Ellis classification, based on the extent to which $T^{ab}$ can be diagonalized by using Lorentz transformations, allows us to classify any SET into one of the four types:\cite{HyE}
\begin{itemlist}
 \item {\it Type I}: the SET can be diagonalized completely, that is there is a basis where $f_i=0$.
 Almost all observed fields belong to this kind.
 \item {\it Type II}: it has a double null eigenvector, which implies that in some basis $f_1=f$, $f_2=f_3=0$, $\rho=\mu+f$, and $p_1=-\mu+f$. 
 Radiation traveling in a null direction has associated a tensor of this form.
 \item {\it Type III}: the SET has a triple null eigenvector, i.~e.~one can take $f_1=f_2=f$, $f_3=0$, $p_1=-\rho+f$, and $p_2=-\rho-f$.
 (See Ref.~\citenum{nonlinear} for this particular expression.)
 \item {\it Type IV}: it has no causal eigenvectors, $f_1=f$, $f_2=f_3=0$, and $p_1=-\rho$.\cite{nonlinear}\\
 (No known classical fields have type III or type IV SETs. 
      The occurrence of type III or type IV SETs seems an intrinsically quantum phenomenon.)
\end{itemlist}
Moreover, for the semi-classical ECs, we take the characteristic 4-velocity to be comoving with respect to the basis in which the
SET can be maximally diagonalized, this is $U_{\widehat{a}}=(-1,\,\vec 0)$.
The different ECs can now be expressed as:
\vspace{2mm}

\noindent WEC:\qquad $\rho- 2\sum_i \beta_i f_i  +\sum_i p_i\beta_i^2\geq 0$.\\
Type I:\,\, $\rho+p_i\geq0$,\,\, $\rho\geq0$.\qquad Type II:\,\,
$\mu+p_2\geq0$,\,\, $\mu+p_3\geq0$,\,\,  $f\geq0$, $\mu\geq0$.\\
Types III and IV cannot satisfy WEC.
\vspace{2mm}

\noindent QWEC:\qquad  $\rho- 2\sum_i \beta_i f_i  +\sum_i p_i\beta_i^2\geq -\zeta (\hbar N/L^4)$.
\vspace{-2mm}
\begin{itemlist}
\item Type I: $\rho+p_i\geq-(\hbar N/L^4)$, $\rho\geq-(\hbar N/L^4)$.
              \item Type II: $p_2\geq- \zeta(\hbar N/L^4)$, $p_3\geq-\zeta(\hbar N/L^4)$, $f\geq-\zeta(\hbar N/L^4)$, $\mu\geq-\zeta(\hbar N/L^4)$.
              \item Type III: $\rho\geq-\zeta(\hbar N/L^4)$, $p_3\geq-\zeta(\hbar N/L^4)$, $|f|\leq\zeta(\hbar N/L^4)$.
              \item Type IV: $\rho\geq-\zeta(\hbar N/L^4)$, $p_2\geq-\zeta(\hbar N/L^4)$, $p_3\geq-\zeta(\hbar N/L^4)$, $|f|\leq\zeta(\hbar N/L^4)$.
\end{itemlist}
\noindent FEC:\qquad $\big[\rho-\vec\beta\cdot\vec f\;\big]^2  - {\textstyle\sum}\left[ f_i-p_i\beta_i\right]^2\geq 0$.\\
Type I:\,\, $\rho^2-p_i^2\geq0$.\qquad 
Type II: \,\,$p_2^2\leq\mu^2$, \,\,$p_3^2\leq\mu^2$, \,\,$\mu f\geq0$.\\
Types III and IV cannot satisfy FEC.

\vspace{2mm}
\noindent QFEC:\qquad $ \big[\rho-\vec\beta\cdot\vec f\;\big]^2  - {\textstyle\sum}\left[ f_i-p_i\beta_i\right]^2 \geq -\zeta\;(\hbar N/L^4)^2$.
\vspace{-2mm}
\begin{itemlist}
 \item Type I: $\rho^2-p_i^2\geq-\zeta\;(\hbar N/L^4)^2$.
              \item Type II: $\mu f\geq -{\zeta \over 4}(\hbar N/L^4)^2$, $\mu^2-p_i^2\geq-\zeta(\hbar N/L^4)^2$.
              \item Type III: $\rho^2-p_3^2\geq-\zeta(\hbar N/L^4)^2$, $|f|\leq\zeta\hbar N/L^4$, $|\rho f|\leq\zeta(\hbar N/L^4)^2$.
              \item Type IV: $|f|\leq \zeta \hbar N/L^4$, $|\rho f|\leq  \zeta(\hbar N/L^4)^2$, 
              $\rho^2-p_{2}^2\geq -\zeta (\hbar N/L^4)^2$,\\
              $\rho^2-p_{3}^2\geq-\zeta(\hbar N/L^4)^2$.
\end{itemlist}

\noindent DETEC:\qquad ${\rm det}\left(T^{ab}\right) = p_1p_2p_3\left[\rho-\left(f_1^2/p_1+f_2^2/p_2+f_3^2/p_3\right)\right] \geq 0.$\\
Type I:\, $\rho\, p_1 p_2 p_3\geq0$.\qquad Type II:\, $-\mu^2 \,p_2 p_3\geq0$.\\
Type III:\, $\rho^3 p_3\geq0$.\qquad\quad Type IV:\, $(\rho^2+f^2) p_2 p_3\leq0$.
\vspace{2mm}

\noindent TOSEC:\qquad $\rho^2 + \sum p_i^2 - 2 \sum f_i^2 \geq 0$.\\
It is trivially satisfied for Type I, II and III.\quad Type IV:\,\,  $2(\rho^2-f^2) + \sum p_i^2\leq0$.

\section{Quantum Vacuum States}
Let us consider the ECs in some quantum vacuum states of special interest.

\subsection{Casimir effect}
The most well known violation of the usual ECs is the Casimir effect, which is a disturbance of the electromagnetic quantum vacuum
due to the presence of two conducting plates. The original calculation of this effect considered two parallel conducting plates separated by
a small distance $a$. Thus, the renormalized vacuum SET is\cite{BD} 
\beq
\langle T^{ab} \rangle =  - {\hbar \pi^2\over720 a^4}\,\, {\rm diag}\left(1,\,-1,\,-1,\,3\right).
\eeq
Unfortunately, the FEC is violated for this vacuum.
As the violations of the ECs are small, the QWEC and QFEC are satisfied taking $L=a$.
The DETEC is also satisfied, with the TOSEC being trivially satisfied.

\subsection{Vacuum polarization in Schwarzschild spacetime}
In the case of the renormalized vacuum expectation value of the SET in Schwarzschild spacetime, the results
depend on which vacuum (associated to what particular class of observers) we consider.

\subsubsection{Hartle--Hawking vacuum}
The Hartle--Hawking vacuum is regular on the past and future horizon and describes a thermal bath at spatial infinity.
On general grounds\cite{Page} we know that the principal energy density and pressures 
characterizing this type I SET are
of order unity. Thus, violations of the ECs (if any) are bounded.\cite{QFEC,nonlinear}

For the particular case of a conformally coupled scalar field, defining $z=2M/r$, we have Page's analytic approximation\cite{Page,HH}
\beq\{\rho\}_{HH}(z)=3\,p_\infty\left[1+2\,z+3\,z^2+4\,z^3+5\,z^4+6\,z^5-33\,z^6\right],
\eeq
\beq\{p_r\}_{HH}(z)=p_\infty\left[1+2\,z+3\,z^2+4\,z^3+5\,z^4+6\,z^5+15\,z^6\right],
\eeq
\beq\{p_{t}\}_{HH}(z)=p_\infty\left[1+2\,z+3\,z^2+4\,z^3+5\,z^4+6\,z^5-9\,z^6\right],
\eeq
with
\beq 
p_\infty={\hslash\over 90(16\pi)^2(2M)^4}\ll\frac{\hslash}{L^4}.
\eeq
Therefore, although WEC, FEC, and DETEC are violated, QWEC, QFEC, and QDETEC are satisfied considering that the black hole horizon sets the characteristic scale of the system.\cite{nonlinear}

\subsubsection{Boulware vacuum}
The Boulware vacuum reduces to Minkowski vacuum at $r\rightarrow\infty$. The Brown--Ottewill approximation allows us
to relate the renormalized SET in this vacuum with that of the Hartle--Hawking one through\cite{BO}
\beq
\langle T^{ab} \rangle_{B} = -{p_\infty\over(1-z)^2}  \,\,{\rm diag}\left(3,\,1,\,1,\,1\right)
+ 
\langle T^{ab} \rangle_{HH}.
\eeq
Thus, although the WEC is unboundedly violated close to the horizon, (implying the QWEC is therefore 
violated), any possible violation
of the FEC would be bounded.\cite{QFEC} The same can be said about the DETEC.\cite{nonlinear}

In the case of a conformally coupled scalar field, combining Page's\cite{Page} and Brown-Ottewill's\cite{BO} analytic approximation,
one obtains\cite{B}
\beq
\{\rho\}_B(z)=-3\,p_\infty \, z^6\;{40-72\,z+33\,z^2\over \left(1-z\right)^2},
\eeq
\beq
\{p_r\}_B(z)=p_\infty \,z^6\;{8-24\,z+15\,z^2\over \left(1-z\right)^2},\qquad
\{p_{t}\}_B(z)=-p_\infty \,z^6\;{\left(4-3\,z\right)^2 \over \left(1-z\right)^2}.
\eeq
Thus, the FEC is satisfied without the need of considering any quantum extension. Furthermore, although the DETEC is violated, 
the QDETEC is satisfied.\cite{nonlinear}

\subsubsection{Unruh vacuum}
This vacuum (associated to an in-falling observer) is regular only on the future horizon and 
describes an outgoing flux of radiation.
As the SET is not diagonal in the Schwarzschild basis, it is not clear what kind of SET it
is in general.\cite{nonlinear}
For a conformally coupled scalar field Visser's semi-analytic approximation\cite{U} 
leads to
\beq
\{\rho\}_U=p_\infty\, {z^2\over 1-z}\;\left(5.349+26.56\,z^2-105.2\,z^3+35.09\,z^4+32.91\,z^5\right),
\eeq
\beq
\{p_r\}_U=p_\infty\, {z^2\over 1-z}\;\left(5.349-26.56\,z^2+65.92z^3-63.37\,z^4+13.32\,z^5\right),
\eeq
\beq
\{p_{t}\}_U=p_\infty\, z^4\;\left(26.56-59.02\,z+38.21\,z^2\right),
\eeq
\beq
\{f\}_U=5.349\,p_\infty \; {z^2\over 1-z}.
\eeq
The only conditions satisfied are QFEC, QDETEC, and QTOSEC. The TOSEC can be violated because the SET
is type IV in the asymptotic region.\cite{nonlinear}

\section{Further comments}

The FEC is widely satisfied in
classical situations and the QFEC in semi-classical scenarios, giving a preferred status to this condition.
However, one can check that the DETEC has to be violated in cosmological scenarios (assuming
general relativity) during the early inflationary and in the current accelerating phase. 

Regarding the applications of this conditions, the QECs have been used in the literature\cite{Bouhmadi-Lopez:2014cca,Bouhmadi-Lopez:2014gza} 
to quantify the degree of violation of the ECs.
The FEC has already been used to prove entropy bounds.\cite{FEC} Nevertheless,
one should not expect to be able to prove the increase area law of black
holes with a condition satisfied by Hawking radiation.
On the other hand, the non-linearities make more difficult the use of the 
nonlinear ECs to prove general features.

\section*{Acknowledgments}
PMM acknowledges financial support from the (Spanish) MINECO
through the postdoctoral training contract FPDI-2013-16161, and the project FIS2014-52837-P.

\end{document}